\documentclass[12pt]{emulateapj}
\slugcomment{Accepted, to appear in ApJ, v735, 2011 July}
\pdfoutput=1
\usepackage{natbib}
\newcommand{\Msun}{\>{\rm M_{\odot}}}

\newcommand{\gtorder}{\mathrel{\raise.3ex\hbox{$>$}\mkern-14mu
             \lower0.6ex\hbox{$\sim$}}}
\newcommand{\ltorder}{\mathrel{\raise.3ex\hbox{$<$}\mkern-14mu
             \lower0.6ex\hbox{$\sim$}}}

\newcommand\eg{{\it e.g.,~}}
\newcommand\ie{{\it i.e.,~}}

\begin{document}

\title{Low-Frequency Oscillations in Global Simulations of Black Hole Accretion}
\author{Sean M. O'Neill\altaffilmark{1,2}, Christopher S. Reynolds\altaffilmark{2,3}, M. Coleman Miller\altaffilmark{2,3}, Kareem A. Sorathia\altaffilmark{2,4}}
\altaffiltext{1}{JILA, University of Colorado, 440 UCB, Boulder, CO 80309}
\altaffiltext{2}{University of Maryland, Department of Astronomy and Maryland Astronomy Center for Theory and Computation, College Park, MD 20742}
\altaffiltext{3}{Joint Space Science Institute (JSI), University of Maryland, College Park, MD 20742}
\altaffiltext{4}{University of Maryland, Department of Mathematics, College Park, MD 20742}

\begin{abstract}
We have identified the presence of large-scale, low-frequency dynamo cycles in a long-duration, global, magnetohydrodynamic (MHD) simulation of black hole accretion. 
Such cycles had been seen previously in local shearing box simulations, but we discuss their evolution over 1,500 inner disk orbits of a global $\pi/4$ disk wedge spanning two orders of magnitude in radius and seven scale heights in elevation above/below the disk midplane. 
The observed cycles manifest themselves as oscillations in azimuthal magnetic field occupying a region that extends into a low-density corona several scale heights above the disk. 
The cycle frequencies are ten to twenty times lower than the local orbital frequency, making them potentially interesting sources of low-frequency variability when scaled to real astrophysical systems. 
Furthermore, power spectra derived from the full time series reveal that the cycles manifest themselves at discrete, narrow-band frequencies that often share power across broad radial ranges. 
We explore possible connections between these simulated cycles and observed low-frequency quasi-periodic oscillations (LFQPOs) in galactic black hole binary systems, finding that dynamo cycles have the appropriate frequencies and are located in a spatial region associated with X-ray emission in real systems.
Derived observational proxies, however, fail to feature peaks with RMS amplitudes comparable to LFQPO observations, suggesting that further theoretical work and more sophisticated simulations will be required to form a complete theory of dynamo-driven LFQPOs.
Nonetheless, this work clearly illustrates that global MHD dynamos exhibit quasi-periodic behavior on timescales much longer than those derived from test particle considerations.
\end{abstract}

\maketitle

\section{Introduction}

The standard physical model of black hole disk accretion accounts for the transport of angular momentum through correlations in magnetohydrodynamic (MHD) disk turbulence driven by the magnetorotational instability (MRI, \citealt{1991ApJ...376..214B}). 
While the linear behavior of this instability is analytically tractable and some analytic progress has been made in evaluating its saturation behaviors (\eg \citealt{2009MNRAS.394..715L,2009ApJ...696.1021V,2009ApJ...698L..72P,2010ApJ...716.1012P}), numerical studies are crucial for understanding the nonlinear evolution of the MRI. 
Local (\ie shearing box) simulations of weakly magnetized accretion, first conducted by \citet{1995ApJ...440..742H}, have been instrumental in facilitating numerical studies of the MRI by permitting smaller domains and larger timesteps at a given resolution than their global counterparts.
Global disk simulations, on the other hand, have been invaluable in elucidating the macroscopic aspects of accretion since they incorporate more natural boundary conditions and allow large-scale conservation behaviors and development of radial structure. 
A crucial question is in what regimes do local simulations serve as a true microcosm for global disk behaviors.
If, for example, large-scale magnetic structures are important in disk coronae, as has been suggested  by \citet{2009ApJ...704L.113B} and \citet{2009ApJ...707..428B}, one might worry that real accretion disks are intrinsically non-local.
Likewise, \citet{2010ApJ...712.1241S} have shown that global magnetic linkages are important even though the evolution of fluid stresses in subdomains of a global simulation are well represented by local simulations with net magnetic flux.
So while the character of MRI turbulence in global simulations appears to be correctly captured by local simulations that include stratification and net magnetic flux, it remains unclear how behaviors involving the development of large-scale field correlations will translate from one simulation regime to the other.

We report in this paper on a long-duration global simulation of black hole accretion that confirms the existence of an interesting phenomenon that previously had only been seen in local disk simulations.
Specifically, we have detected prominent low-frequency ``dynamo cycles'' in the azimuthal magnetic field evolution of a simulated global accretion disk.
Dynamo cycles are commonly invoked as the explanation for the observed 22-year solar magnetic cycle (see \eg \citealt{1955ApJ...122..293P}, \citealt{1961ApJ...133..572B}, \citealt{1969ApJ...156....1L} or the review by \citealt{1985ARA&A..23..379B}).
More directly relevant to our work, however, is the fact that such cycles have appeared in many shearing box simulations of accretion disks (\citealt{1995ApJ...446..741B,1996ApJ...463..656S,2000ApJ...534..398M,2004ApJ...605L..45T,2006ApJ...640..901H,2009ApJ...697.1269J,2009ApJ...691L..49S,2010ApJ...708.1716S,2010MNRAS.405...41G,2010ApJ...713...52D,2010arXiv1010.0005S}).
These simulated cycles are typically seen to have periods on the order of tens of local orbital periods, with the exact number varying somewhat with the details of the simulation (the vertical domain, in particular, was a limiting factor in many of the earliest simulations).
As  \citet{2010ApJ...708.1716S}, \citet{2010MNRAS.405...41G}, and \citet{2010ApJ...713...52D} discuss, the generic ``butterfly'' pattern can be attributed to the vertical rising of azimuthal field due to the combined effects of dynamo action near the disk midplane and the Parker instability at higher elevation.
Until now, however, such features have not been reported in global simulations.

In analogy to the results of local simulations, our global disk simulation produces dynamo cycles with oscillation frequencies that are ten to twenty times lower than the local orbital frequencies for a large range of radii.
Furthermore, the radial extent of our simulation captures the sharing of dynamo power at peak frequencies across relatively large radial ranges.
As such, dynamo cycles provide a tantalizing source of variability at frequencies comparable to astronomically observed low-frequency quasi-periodic oscillations (LFQPOs) in multiple galactic black hole binary candidates.

We proceed with a discussion of our numerical model ($\S 2$), followed by a detailed description of the resulting dynamo cycles ($\S 3$).
We then compare our global simulation to published local simulations and discuss broader observational implications ($\S 4$), followed by our conclusions ($\S 5$).

\section{Numerical Model}

Our fully three-dimensional magnetohydrodynamic (MHD) simulation employs a modified version of the publicly available ZEUS-MP (version 2) code, described in \citet{1992ApJS...80..753S}, \citet{1992ApJS...80..791S}, and \citet{2006ApJS..165..188H}.
ZEUS-MP uses an Eulerian finite difference scheme accurate to second order in space to solve the equations of ideal compressible MHD,
\begin{equation}
\frac{D\rho}{Dt}=-\rho{\bf\nabla\cdot{v}},
\end{equation}
\begin{equation}
\rho\frac{D{\bf{v}}}{Dt}=-{\bf\nabla}P+\frac{1}{4\pi}({\bf\nabla\times{B}}){\bf\times{B}}-\rho{\bf\nabla}\Phi,
\end{equation}
\begin{equation}
\rho\frac{D}{Dt}\left(\frac{e}{\rho}\right)=-{P}{\bf\nabla\cdot{v}}-\Lambda,
\end{equation}
\begin{equation}
\frac{\partial{\bf{B}}}{\partial{t}}={\bf\nabla\times}({\bf{v}\times{B}}),
\end{equation}
where 
\begin{equation}
\frac{D}{Dt}\equiv\frac{\partial}{\partial{t}}+{\bf{v\cdot\nabla}}.
\end{equation}
We employ a gamma-law ($\gamma=5/3$) gas equation of state.
Timesteps are set by the usual Courant condition, and a protection routine prevents the density and pressure from reaching artificially small and/or negative values.
The only adjustments we have made to the fundamental ZEUS-MP algorithm involve this protection routine and, as described below, the introduction of modified gravity and the gas cooling function, $\Lambda$.

Gravity is modified in our simulation to emulate the relevant effects of general relativity through a pseudo-Newtonian potential \citep{1980A&A....88...23P} of the form:
\begin{equation}
\label{eq:pnpot}
\Phi=-\frac{GM}{R-2r_{\rm g}},\qquad r_{\rm g}\equiv\frac{GM}{c^2}.
\end{equation}
This approach accurately captures for a Schwarzschild spacetime the position of the innermost stable circular orbit (ISCO) at $r = 6 r_{\rm g}$, the period of which is $\tau _{\rm ISCO} \approx 61.6~{\rm GM/c}^3$ in this potential.

\begin{figure*}[t]
\begin{center}
\includegraphics[type=pdf,ext=.pdf,read=.pdf,height=0.27\textwidth]{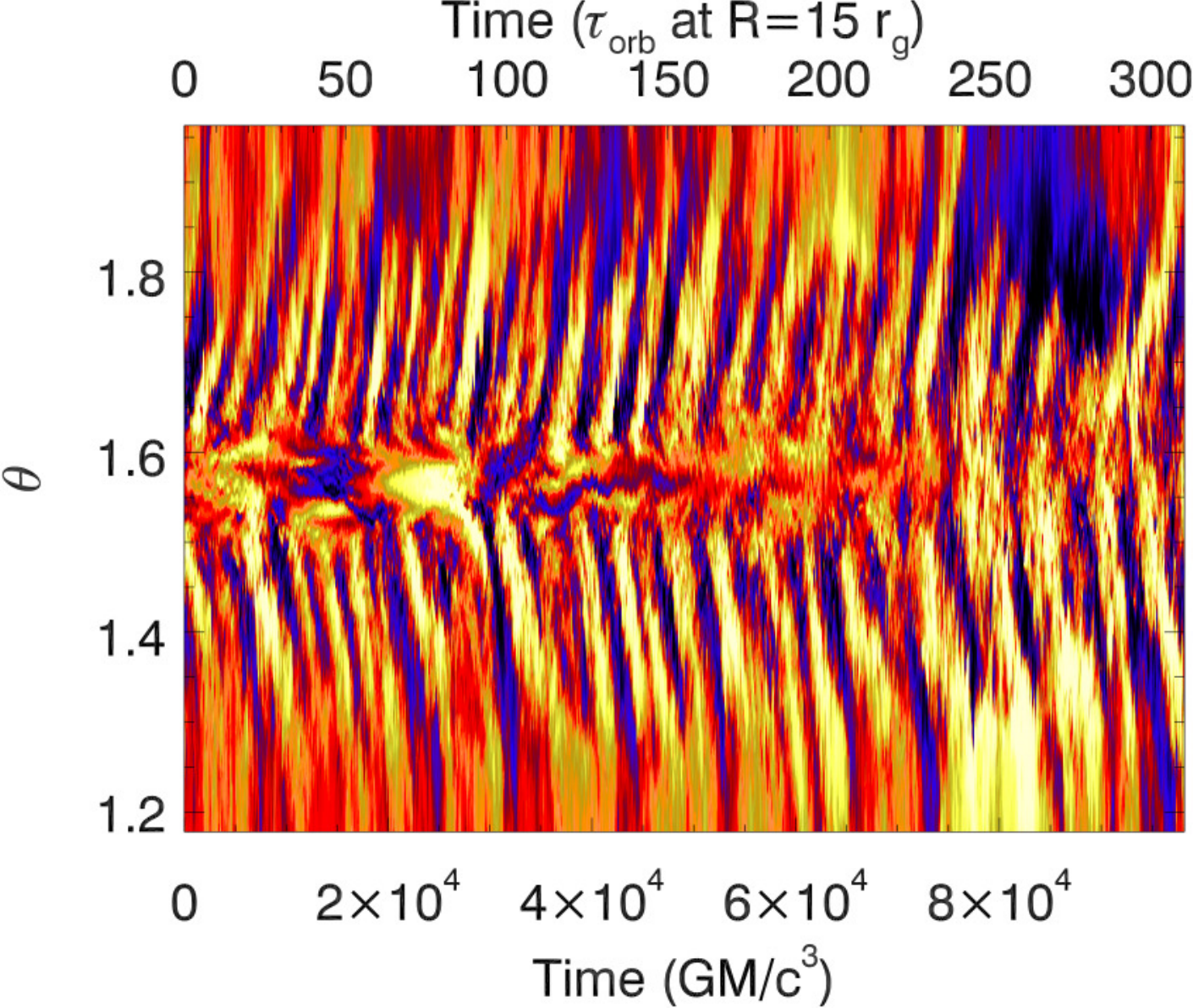}
\includegraphics[type=pdf,ext=.pdf,read=.pdf,height=0.27\textwidth]{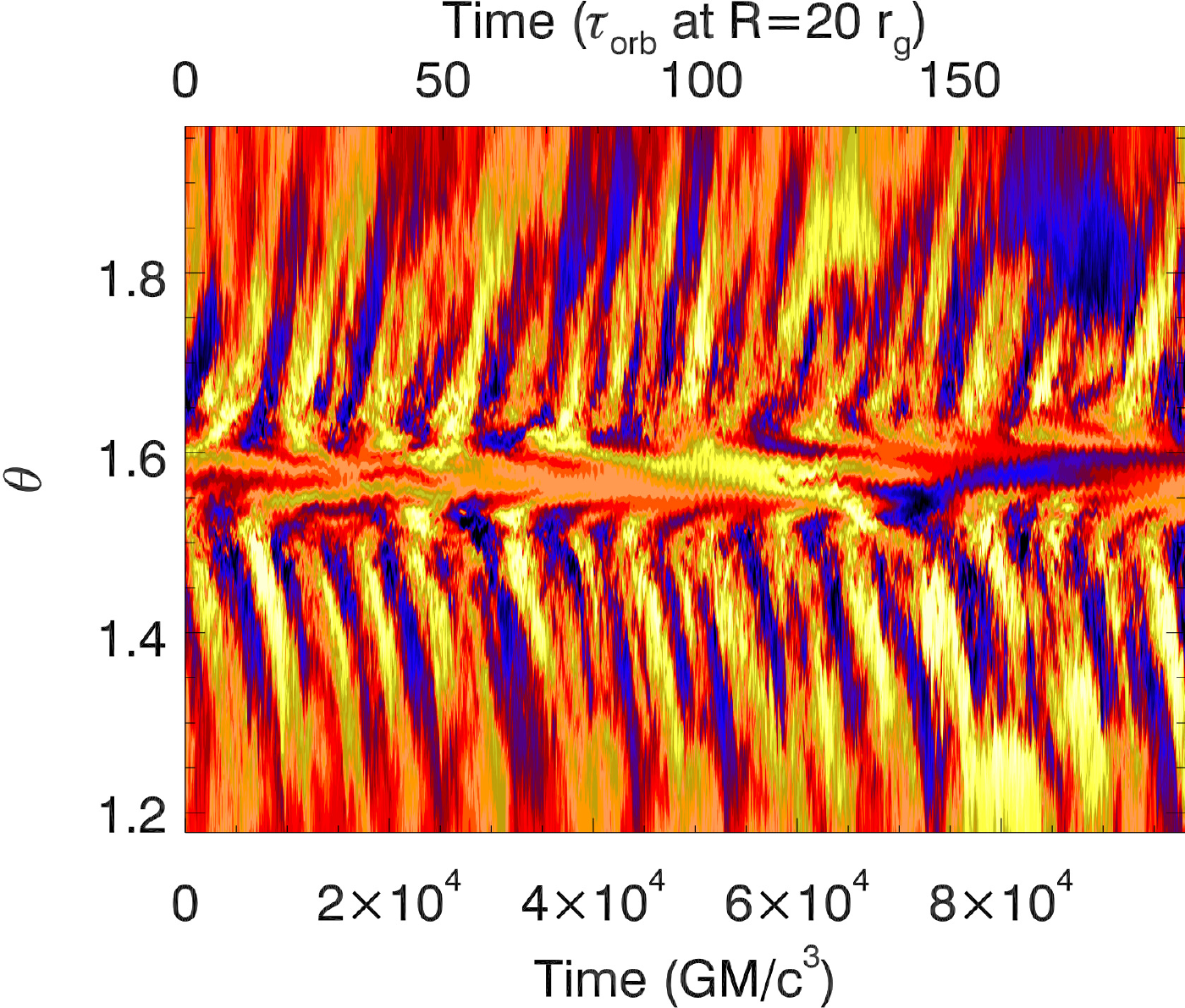}
\includegraphics[type=pdf,ext=.pdf,read=.pdf,height=0.27\textwidth]{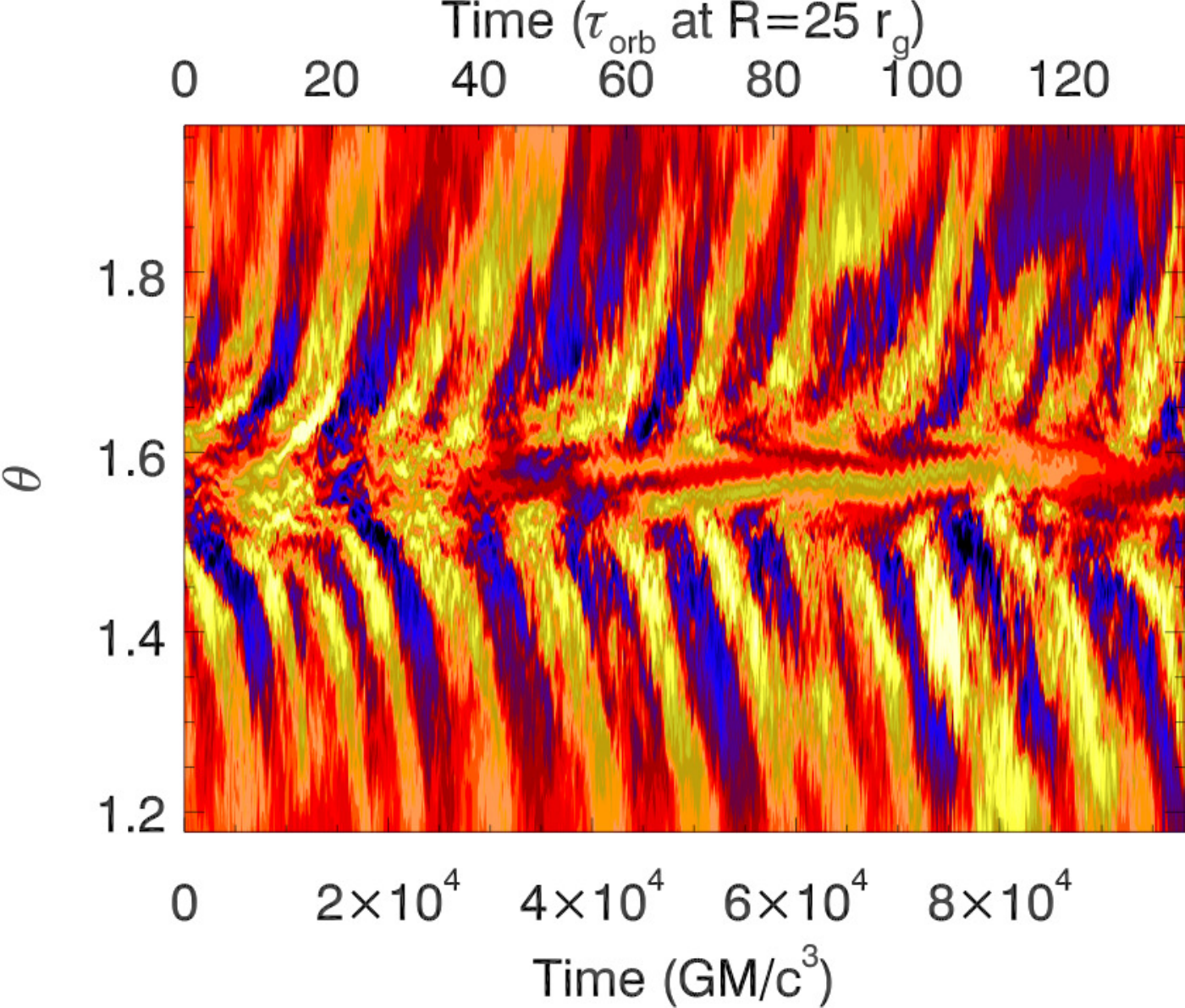}
\caption{Space-time evolution of azimuthal magnetic field at $R = 15~{\rm r_g}$ ({\it left}), $20~{\rm r_g}$ ({\it center}), and $25~{\rm r_g}$ ({\it right}).  The field has been averaged only in the azimuthal direction.  Bright (yellow) regions indicate strong, positive fields while dark (violet) regions are strongly negative.  The disk midplane is located at $\theta = \pi/2$ and $\Delta \theta =0.05$ is one scaleheight.  Time is shown both in units of ${\rm GM/c^3}$ and in terms of the local orbital period, $\tau_{\rm orb}$.  These ``butterfly diagrams'' illustrate that the azimuthal field changes sign over timescales on the order of ten times the local orbital period.}
\end{center}
\end{figure*}

We initialize our computational grid using spherical coordinates ($R,\theta,\phi$) that span $R \in [4 {\rm r_g},400 {\rm r_g}], \theta \in [0.05\pi,0.95\pi], \phi \in [0,\pi/4)$.
The grid is non-uniform, logarithmically increasing in $R$ with a maximum $R$ resolution of $\Delta R = 0.025$ r$_{\rm g}$ at the inner edge of the grid.
The zone aspect ratio is approximately $\Delta R: R\Delta \theta: R\Delta \phi = 3:1:6$ everywhere within seven scale heights above/below the disk midplane (\ie at $\theta = \pi/2$), and each scale height is resolved with $25$ computational zones in this region. 
Outside of this region, the $\theta$ resolution logarithmically increases outward.
The total grid size is $N_R \times N_\theta \times N_\phi = 512 \times 384 \times 64 = 1.26 \times 10^7$ zones.
Standard ZEUS-MP boundary treatments are employed, with a restricted boundary condition that permits outflow only in the $\pm R$ directions.
Reflecting conditions are used near the coordinate pole in $\theta$ (as in \citealt{2003ApJ...592.1060D}, for example), and periodic conditions are applied in $\phi$.

The initial conditions correspond to a thin, axisymmetric disk of constant midplane density and radially decreasing pressure:
\begin{equation}
\rho(R,\theta)=\rho_{\rm 0}\exp\left(-\frac{\cos^2\theta}{2(h/r)^2\sin^2\theta}\right),
\end{equation}
and 
\begin{equation}
p(R,\theta)=\frac{GMR(h/r)^2\sin^2\theta}{(R-2r_{\rm g})^2}~\rho(R,\theta)
\end{equation}
where $\rho_{\rm 0}$ is the initial density in the disk midplane and $h$ is the effective scale height.
The disk aspect ratio is initialized to $h/r=0.05$ everywhere, and a cooling function $\Lambda$ is implemented to maintain this aspect ratio with a cooling time ($\tau_{\rm cool}$) that is related to the local orbital period ($\tau_{\rm orb}$) using an approach similar to that described in \citet{2009ApJ...692..411N}.
The exact form of the cooling function is $\Lambda = f(e-e_{\rm targ})/ \tau_{\rm cool}$, where $f = 0.5 [(e-e_{\rm targ})/|e-e_{\rm targ}|+1]$ is a threshold function that enables cooling only when the internal energy $e$ is greater than the target energy $e_{\rm targ}$ (and which is equal to zero when $e_{\rm targ} > e$).
The target energy is chosen so that $e_{\rm targ} \propto \rho v_{\phi}^2 (h/r)^2$, which comes from the assumption that $c_s \sim (h/r) v_{\phi}$ in thin disks.
Estimating the cooling time as the thermal timescale of the disk, we choose $\tau_{\rm cool}=\tau_{\rm orb}/\alpha = 10 \tau_{\rm orb}$, which corresponds to a \citet{1973A&A....24..337S} alpha disk with $\alpha=0.1$.
While the true effective alpha parameter varies spatially and in time over the evolution of an MHD disk, this choice provides an adequate order-of-magnitude estimate for the implementation of cooling.
Additionally, we ran a short test simulation that revealed that the frequency range of the dynamo cycle signal discussed in the following section was insensitive to the presence or absence of cooling, although the signal itself was more pronounced in the case with cooling.

The initial velocity profile is entirely azimuthal and is set such that the effective centrifugal force balances the gravity of the central object in the disk midplane.
The initial magnetic field is completely poloidal in orientation and consists of a series of magnetic field loops that span several local scale heights in both height and width (\eg see \citealt{2009ApJ...692..869R}).
The average ratio of gas-to-magnetic pressure is initialized to $\beta\approx1000$. 

\section{Results}

The following analysis treats the disk only after it has relaxed away from its initial conditions.
Specifically, we allow the disk to evolve for $150$ ISCO orbits ($>9200$~GM/c$^3$) and define $t=0$ to correspond to the end of this initialization period.
The simulation is then followed post-initialization for more than 1500 ISCO orbits, from $t=0$ to $t\approx9.8\times 10^4$~GM/c$^3$.
During and after the initialization, the disk naturally evolves to a turbulent state as a result of the MRI.
Deferring a detailed discussion of the full disk evolution to a future work, we focus here on an interesting set of coherent behaviors that emerge from this turbulence.

\subsection{Azimuthal Magnetic Field Oscillations}

Figure 1 shows the azimuthal field strength (also azimuthally averaged) as a function of both time and elevation from the midplane for three distinct radii.
At all radii, the disk midplane is located at $\theta=\pi/2$, while $\Delta \theta =0.05$ corresponds to one scaleheight for fixed $h/r$. 
Each panel of the figure shows that the azimuthal field at a given location reverses sign multiple times.
This variation can be seen for several scale heights above and below the disk midplane, forming a ``butterfly'' pattern analogous to that which has been observed in shearing box simulations of accretion disks.
The period of field reversal is generally longer for larger distances from the central object, although there are also some common features seen at all radii.
While it is difficult to tell from these diagrams whether the variability is truly periodic, it appears that the reversals in field orientation take place on timescales on the order of tens of local orbital periods.
Near the maxima of the field cycles, the azimuthal field strength can be twice as strong as the RMS value of the azimuthal field measured over the simulation duration in the same region. 

\begin{figure*}[t]
\begin{center}
\includegraphics[type=pdf,ext=.pdf,read=.pdf,height=0.24\textwidth]{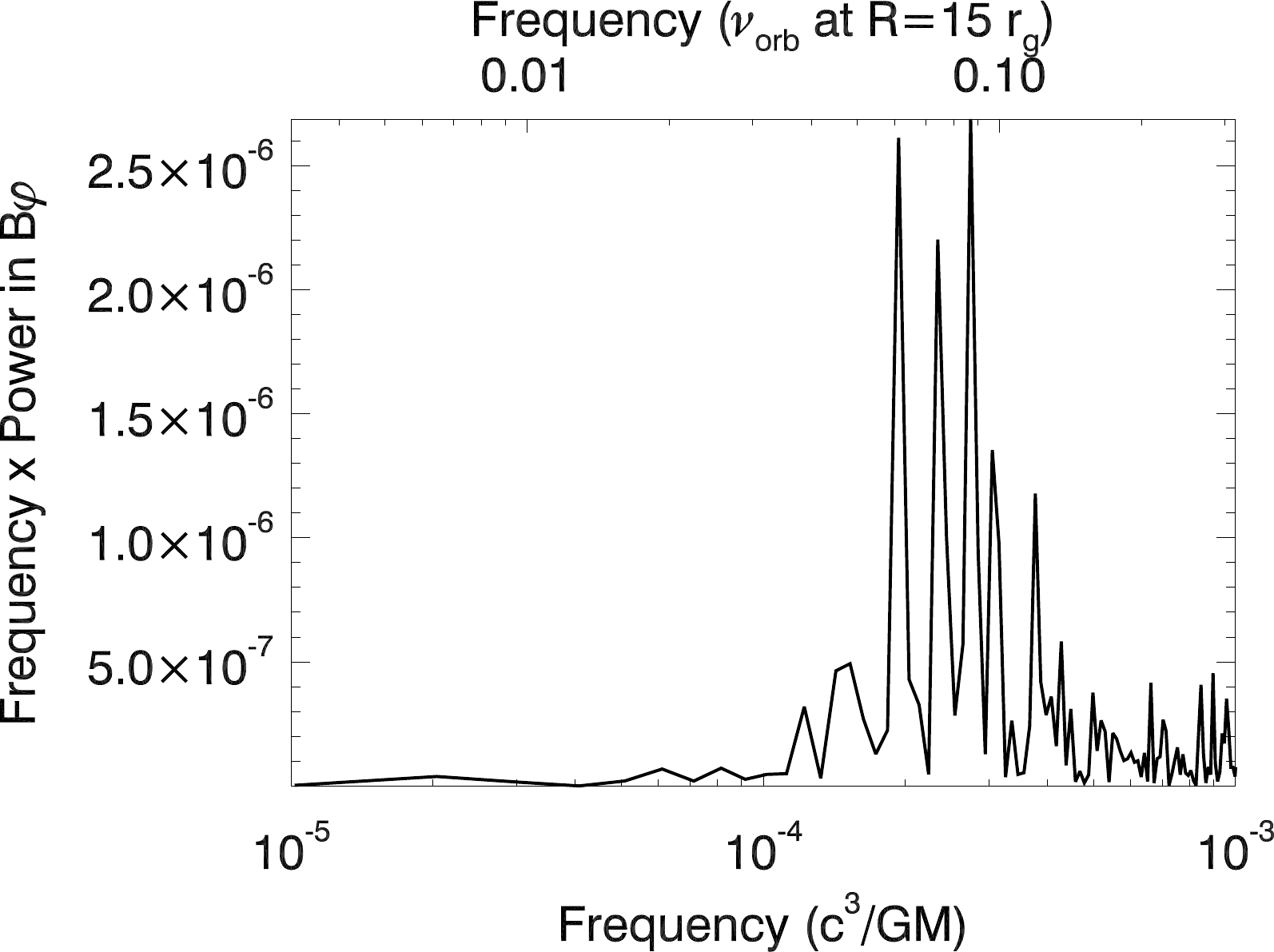}
\includegraphics[type=pdf,ext=.pdf,read=.pdf,height=0.24\textwidth]{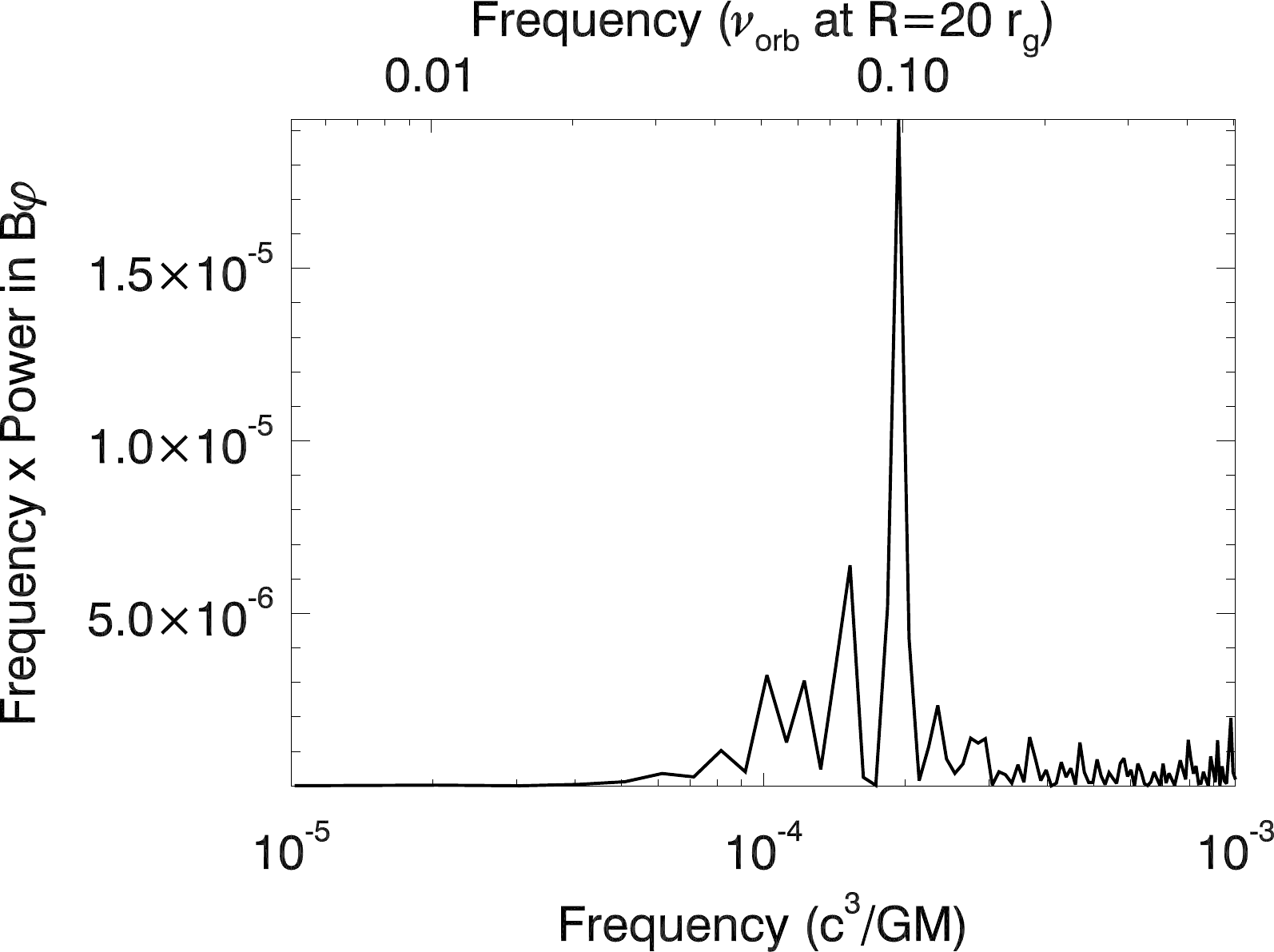}
\includegraphics[type=pdf,ext=.pdf,read=.pdf,height=0.24\textwidth]{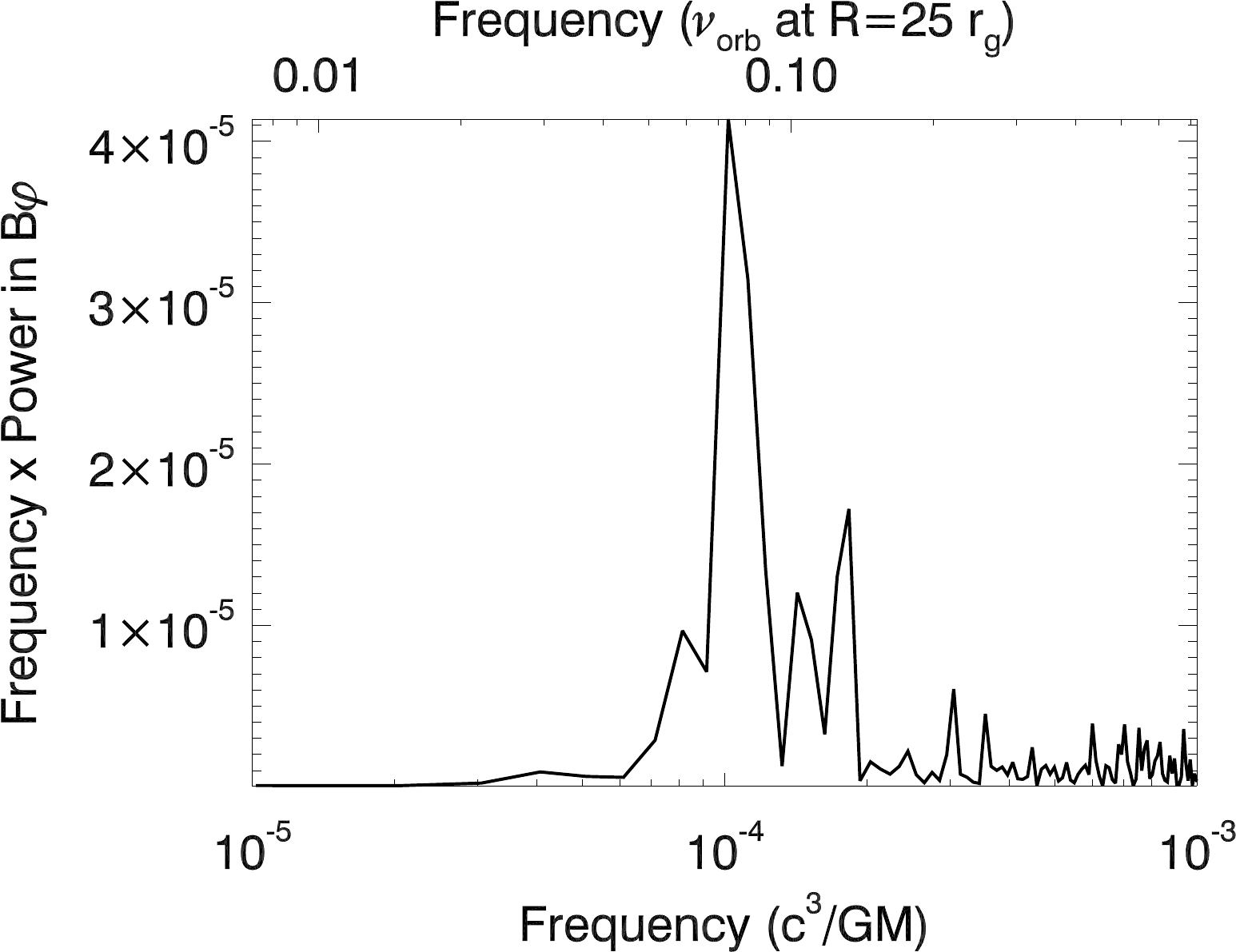}
\caption{Frequency-weighted PSDs ($\nu P$, in arbitrary units) of azimuthal magnetic field strength taken over the duration of the simulation at  at $R = 15~{\rm r_g}$ ({\it left}), $20~{\rm r_g}$ ({\it center}), and $25~{\rm r_g}$ ({\it right}), measured at two scale heights above the disk midplane and averaged both azimuthally and over a radial domain comparable to a scale height.  Each radius features strong, multi-peaked signals corresponding to the oscillations seen in Figure 1.}
\end{center}
\end{figure*}

To explore this variability in more detail, we show in Figure 2 the power spectral density (PSD), defined as $P(\nu)=\eta|\bar{f}(\nu)|^2$, where $\eta$ is a normalization constant and $\bar{f}(\nu)$ is the Fourier transform
\begin{equation}
\bar{f}(\nu)=\int{f(t)e^{-2{\pi}i{\nu}t}dt},
\label{eq:ft}
\end{equation}
of a time series $f(t)$.
In each panel, the azimuthal magnetic field has been azimuthally averaged and summed over a radial range comparable to the local scale height before the PSD is computed for a location two scale heights above the disk midplane.
In this case, the time series is taken to be the duration of the simulation after initialization. 
All three panels show strong power enhancements at frequencies roughly comparable to ten to twenty local orbital periods (\ie $0.05-0.1~\nu_{\rm orb}$, where $\nu_{\rm orb}$ is the local orbital frequency).
It is interesting to note that all radii also show multiple peaks, suggesting that the phenomenon is not always simply related to the local orbital period.
In fact, there is some overlap between adjacent radii, as seen in the shared frequency of the strongest peaks for both $R=15~{\rm r}_g$  and $R=20~{\rm r}_g$.

As a coarse estimate of the significance of these peaks, we can compare their strengths to the mean power as estimated from nearby frequencies.
For a PSD of a single time series, the mean is comparable to the standard deviation of the power distribution (\eg \citealt{1992nrca.book.....P}), implying that the ratio of peak-to-mean power can be used to estimate peak significance.
In the case of $R=15~{\rm r}_g$, for example, the two highest peaks are approximately 15-20 times the mean power as extrapolated from nearby frequency ranges.
The case of $R=20~{\rm r}_g$ is even more convincing as the primary peak is approximately 90 times stronger than the mean while the secondary peak is roughly 40 times above the mean.
Even if our approximations somehow underestimate the mean power by factors of a few, each radius features multiple peaks that stand significantly above the noise.

\begin{figure*}[t]
\begin{center}
\includegraphics[type=pdf,ext=.pdf,read=.pdf,height=0.24\textwidth]{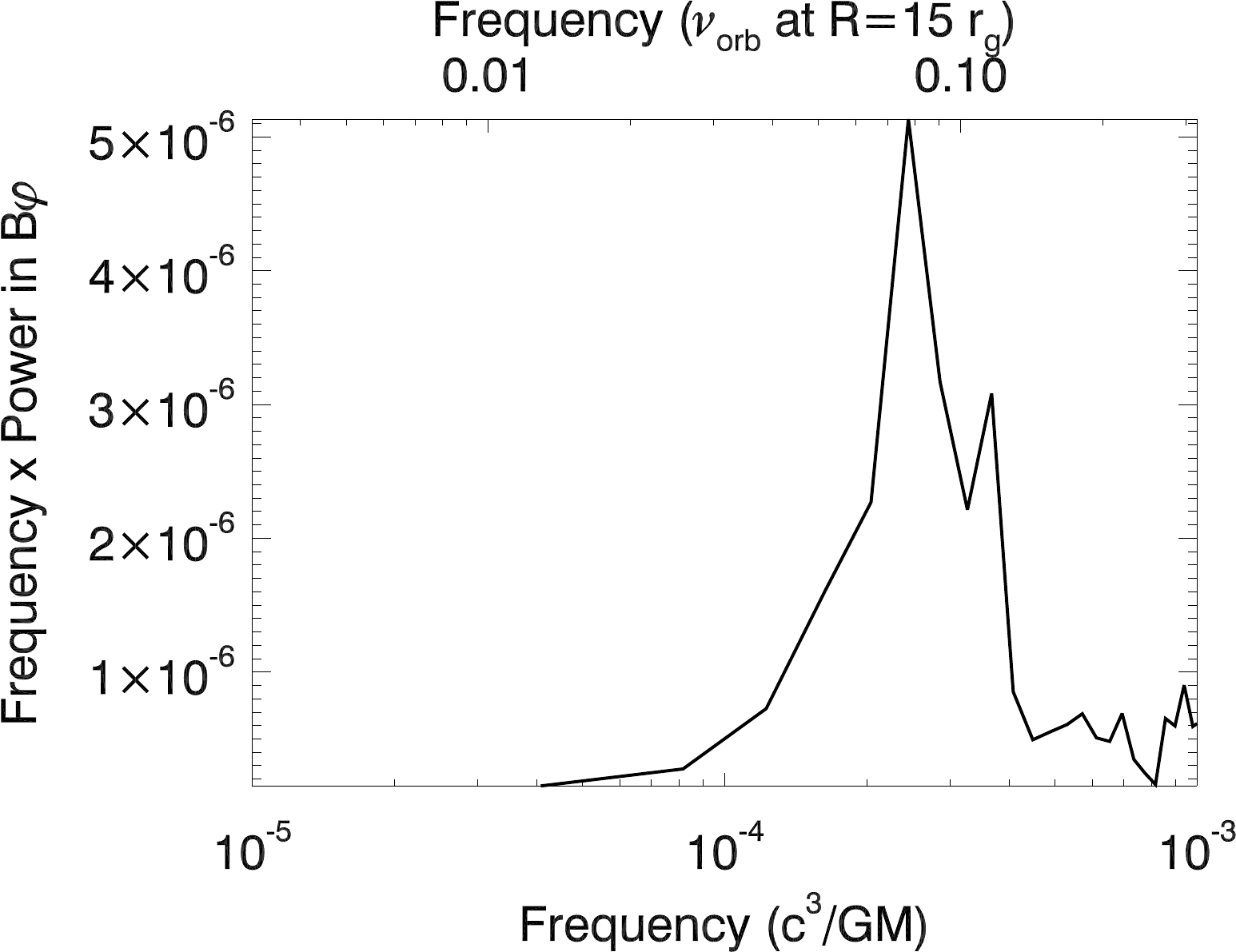}
\includegraphics[type=pdf,ext=.pdf,read=.pdf,height=0.24\textwidth]{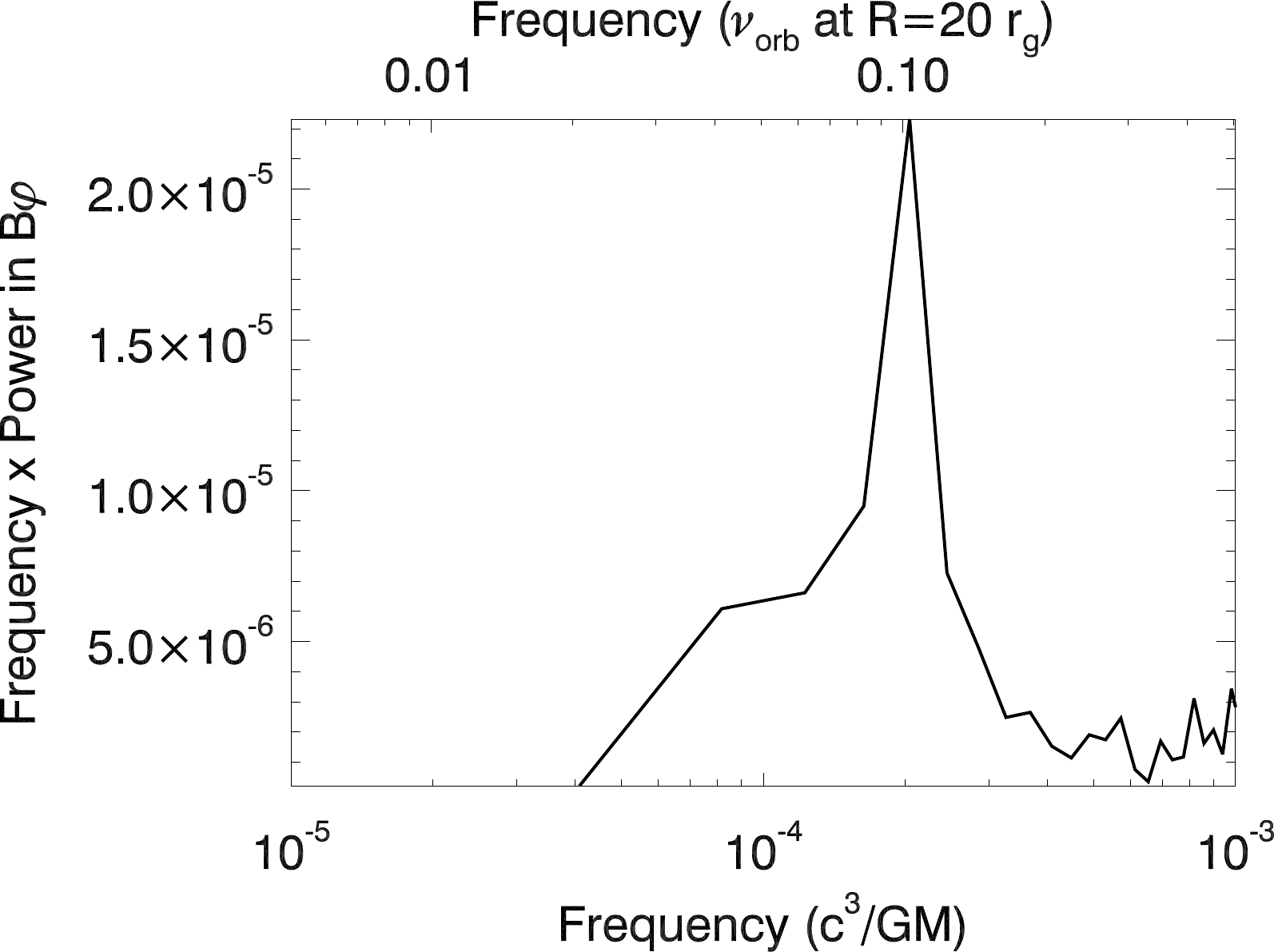}
\includegraphics[type=pdf,ext=.pdf,read=.pdf,height=0.24\textwidth]{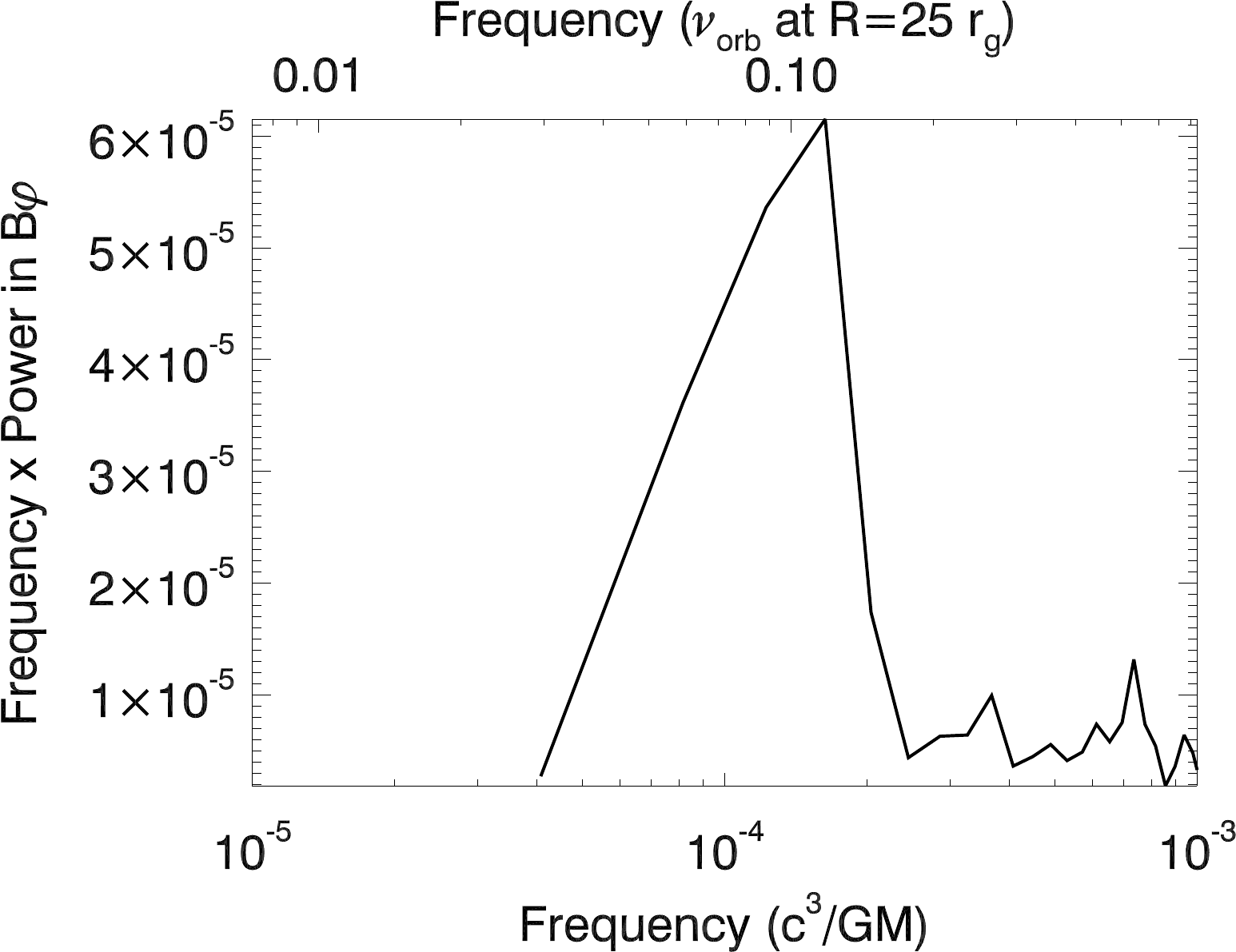}
\caption{Frequency-weighted $\overline{\rm PSD}$s ($\nu \overline{P}$, in arbitrary units) of azimuthal magnetic field strength as in Figure 2, but constructed from averages over four independent time series.  Each radius features strong signals suggesting that the oscillations seen in Figures 1 and 2 are robust.}
\end{center}
\end{figure*}

To further evaluate the significance of these features, we also construct the average power spectral density ($\overline{\rm PSD}$), defined as $\displaystyle \overline{P}(\nu) = (1/N)\sum\limits_{i=1}^N P_i(\nu)$ over a set of $N$ independent time series $f_i(t)$.
This approach has the advantage of reducing the standard deviation of the PSD features at the expense of the available frequency domain and resolution (see \eg \citealt{1989tns..conf...27V,1992nrca.book.....P,2003MNRAS.345.1271V}).
Given the frequencies of interest and the total duration of the simulation, we can afford to take only $N=4$ independent time series, which increases the significance of the averaged PSD by a factor of two over the unaveraged case.
Figure 3 shows the $\overline{\rm PSD}$s of the azimuthal magnetic field over the same regions depicted in Figure 2.
As in Figure 2, we see peaks (now broadened, but at higher significance) that sit approximately ten to twenty times below the local orbital period.
While the peaks are sufficiently broad that they overlap for different radii, the frequency resolution of the $\overline{\rm PSD}$ is insufficient to determine how significant this overlap is.

Figure 4 shows the PSD of the azimuthal magnetic field as a function of both frequency and radius, so that we are better able to evaluate the radial dependence of the power profile.
As in Figure 2, the PSD has been computed over the entire time series.
The vertical features in Figure 4 illustrate that power is shared at distinct frequencies across large radial intervals, with single peaks often stretching across radial ranges of 10 $r_{\rm g}$ or more.
Taken in aggregate, however, the power distribution reflects the radial run of the orbital frequency.
Specifically, the power is bounded by $\sim \nu_{\rm orb}/6$ on the high-frequency end and $\sim \nu_{\rm orb}/30$ on the low.
So even though a given power peak may radially span multiple orbital frequencies, the range of peak frequencies remains approximately proportional to the local orbital frequency.
This pattern only stands out clearly from the noise for $R \gtorder 10$ r$_{\rm g}$.
Inward of this region, the broadband noise associated with accretion across the ISCO masks any obvious trend.

\begin{figure}[t]
\begin{center}
\includegraphics[type=pdf,ext=.pdf,read=.pdf,width=0.45\textwidth]{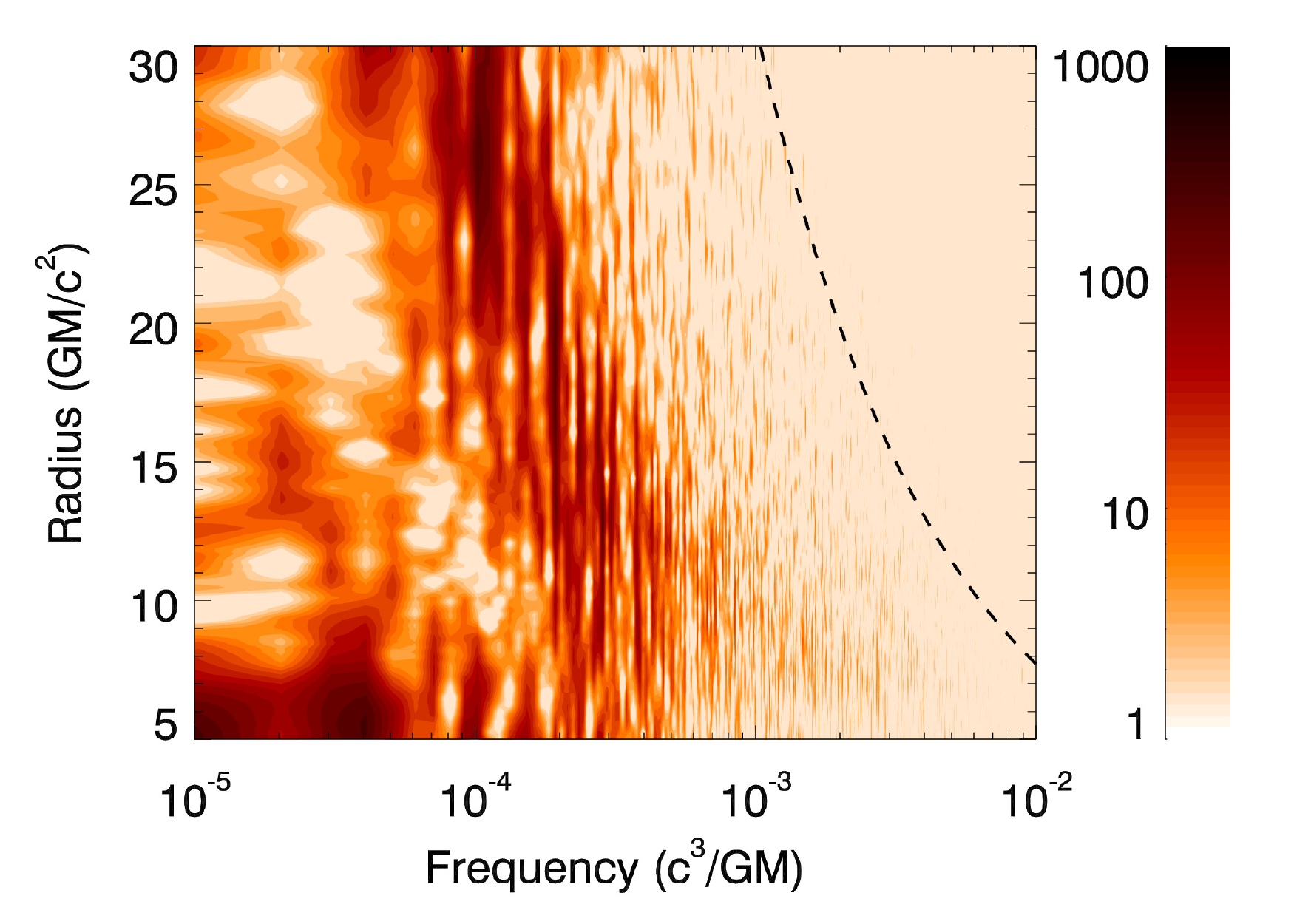}
\caption{PSD ($P$, in arbitrary units) of azimuthal field strength over a range of radii, as measured at two scale heights above the disk midplane and averaged azimuthally.  Also shown as a dashed line is the local orbital frequency.  That the dark (red) band of enhanced power runs parallel to the dashed line demonstrates that the {\it range} of oscillation frequencies is roughly proportional to the local orbital frequency.  A given power peak, however, can often be seen to stretch across a large radial range.}
\end{center}
\end{figure}

\begin{figure*}[t]
\begin{center}
\includegraphics[type=pdf,ext=.pdf,read=.pdf,height=0.24\textwidth]{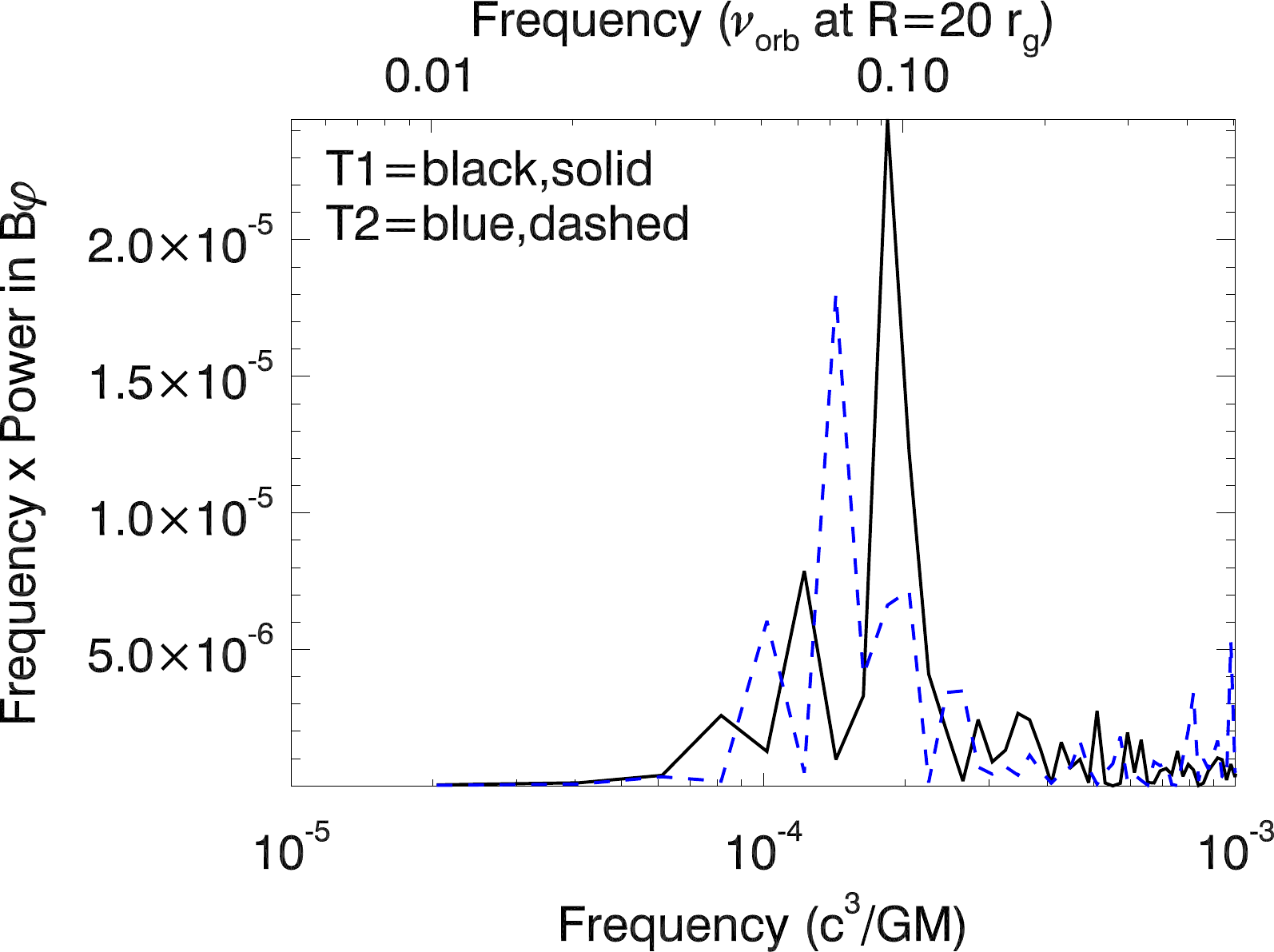}
\includegraphics[type=pdf,ext=.pdf,read=.pdf,height=0.24\textwidth]{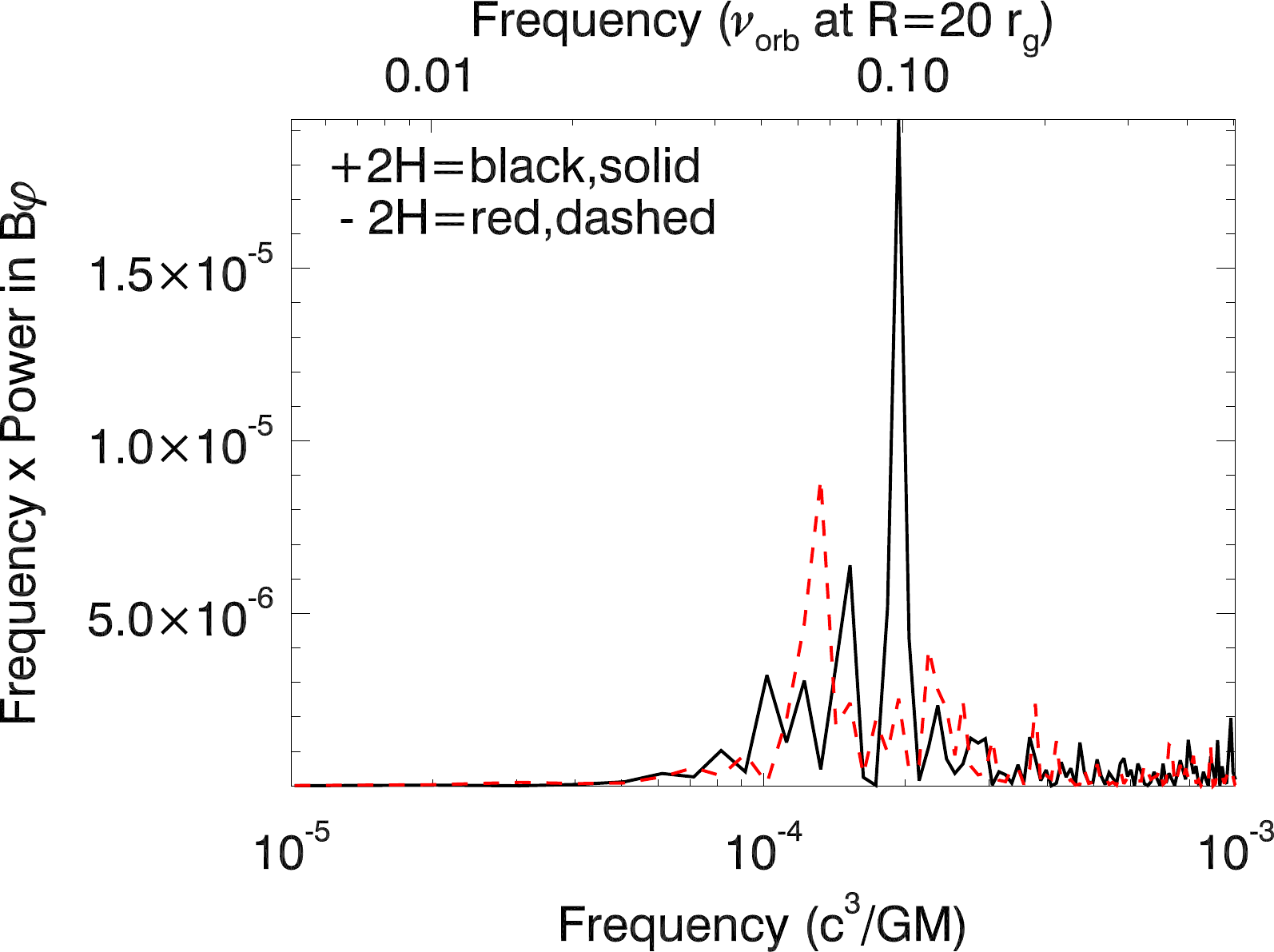}
\includegraphics[type=pdf,ext=.pdf,read=.pdf,height=0.24\textwidth]{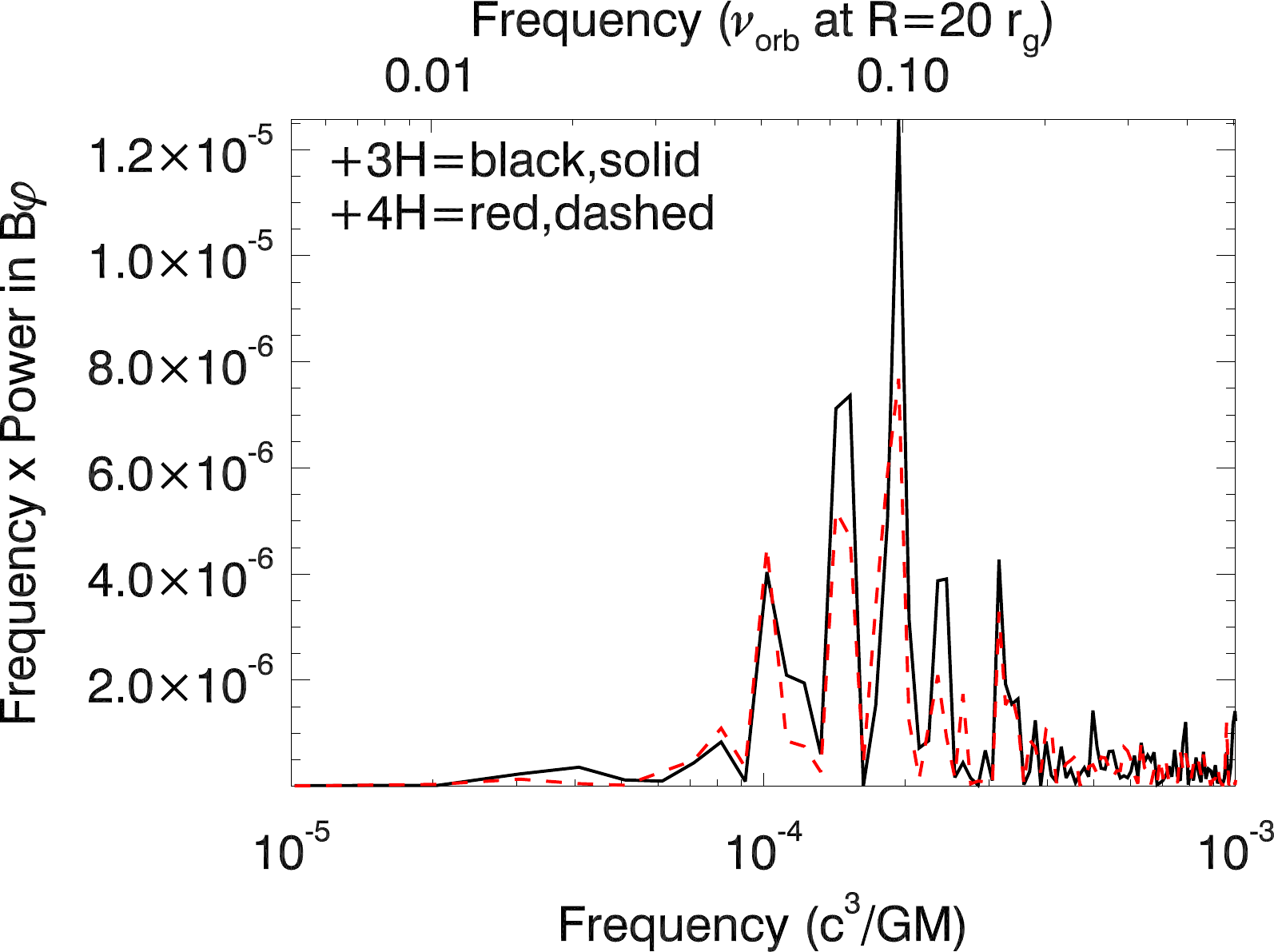}
\caption{Frequency-weighted PSDs ($\nu P$, in arbitrary units) of azimuthal magnetic field strength measured at $R=20~r_{\rm g}$.  The left panel illustrates the time variation of the signal by comparing PSDs taken over two independent halves of the total simulation time (after startup).  The middle panel shows PSDs corresponding to two distinct regions (two scale heights) above and below the disk midplane.  The right panel shows PSDs for three and four scale heights above the midplane, which are similar in frequency to one another and to the signal at two scale heights.}
\end{center}
\end{figure*}

To further elucidate the nature of this variability, Figure 5 presents analyses of the azimuthal field at $R=20~{\rm r_g}$ after various cuts and segregations have been imposed.
In the left panel of Figure 5, we show the PSDs for this region when it is divided into two independent time series (T1 represents the first half of the simulation, T2 the second), each of which is approximately $5 \times 10^4$ GM/c$^3$ in duration.
Note that the peaks are not constant in frequency, suggesting that the multiple peaks in Figures 2-4 are at least partly caused by frequencies migrating in time.
It is tempting to claim that the peaks move from higher to lower frequencies over time, but it is challenging in practice to follow individual peaks without a much longer time baseline.

The middle panel of Figure 5 shows the PSD for the full time series, now comparing the regions above and below the midplane.
These peaks, too, fail to perfectly align even though the disk starts from an approximately symmetric state.
This is perhaps not surprising given that features in the disk turbulence are also seen to evolve asymmetrically about the midplane.
That said, this top-bottom asymmetry suggests that the exact frequency of a single field oscillation may be less useful as a diagnostic tool than the range of frequencies observed.
The rightmost panel of Figure 5 compares the PSDs for regions three and four scale heights above the disk midplane.
The similarity of these peaks to one another and to the locations (if not the relative amplitudes) of the analogous peaks in Figure 2 suggests that, as expected, the variability on a given side of the disk features similar frequencies at different heights from the disk midplane.

\subsection{Observational Proxies}

Thus far, we have focused only on the behavior of the azimuthal magnetic field.
We would also like to explore whether any derived quantities and/or observational proxies oscillate in a similar manner.
Figure 6 shows PSDs of three derived quantities: the integrated $R-\phi$ stresses
\begin{equation}
W_{R,\phi}={\int{(\rho{v_{\rm{R}}v_\phi}-B_{\rm{R}}B_{\rm{\phi}}/4\pi)}dV},
\end{equation}
the integrated Ohmic dissipation
\begin{equation}
P_{\rm{Ohm}}=\int\frac{(\nabla\times{B})^2}{\sigma}dV
\end{equation}
(where $\sigma$ is assumed constant), and the mass accretion rate
\begin{equation}
\dot{M}=-\int_{\rm ISCO}\rho{v_{\rm{R}}dA_{\rm{R}}}.
\end{equation}
Integration ranges are provided in the caption to Figure 6, and the RMS-normalized PSD for each quantity is computed over the entire simulated time series. 
Of these three quantities, only the total stresses feature oscillations that stand above the local noise at frequencies comparable to those at which the azimuthal field oscillates.
This is not completely surprising since the dominant Maxwell stresses depend upon the azimuthal field, but it is nonetheless interesting that a quantity spatially integrated over a large radial range, five scale heights, and the entire azimuthal range of the grid still selects a set of distinct frequencies.
In contrast, both the Ohmic dissipation and accretion rate are dominated by noise, particularly at low frequencies.
In practice, we would have to convincingly model and subtract what appears to be a red (\ie Brownian) noise spectrum in the full data set to prove that any features in the full PSDs were significant.

\begin{figure*}[t]
\begin{center}
\includegraphics[type=pdf,ext=.pdf,read=.pdf,height=0.2\textwidth]{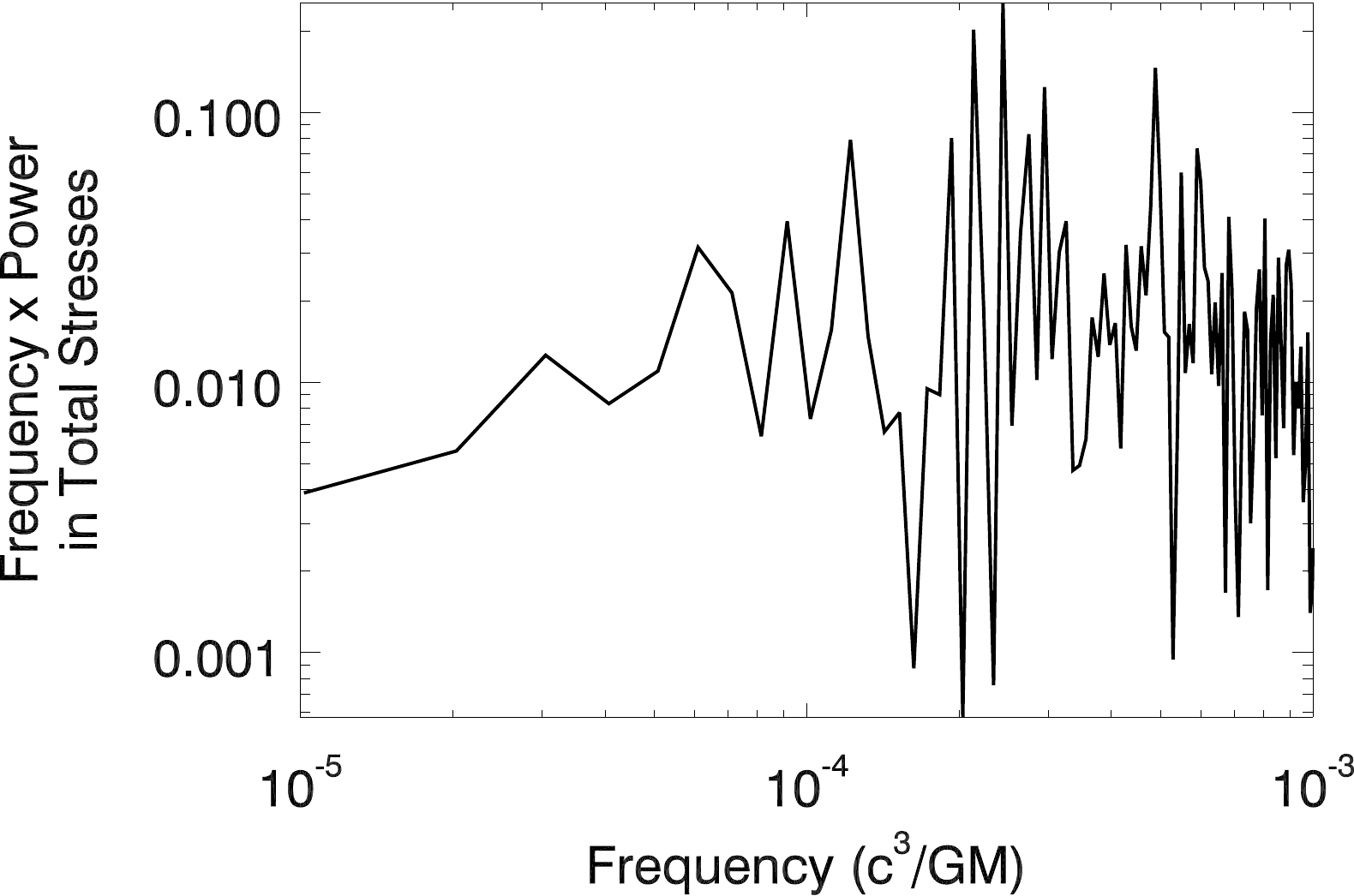}
\includegraphics[type=pdf,ext=.pdf,read=.pdf,height=0.2\textwidth]{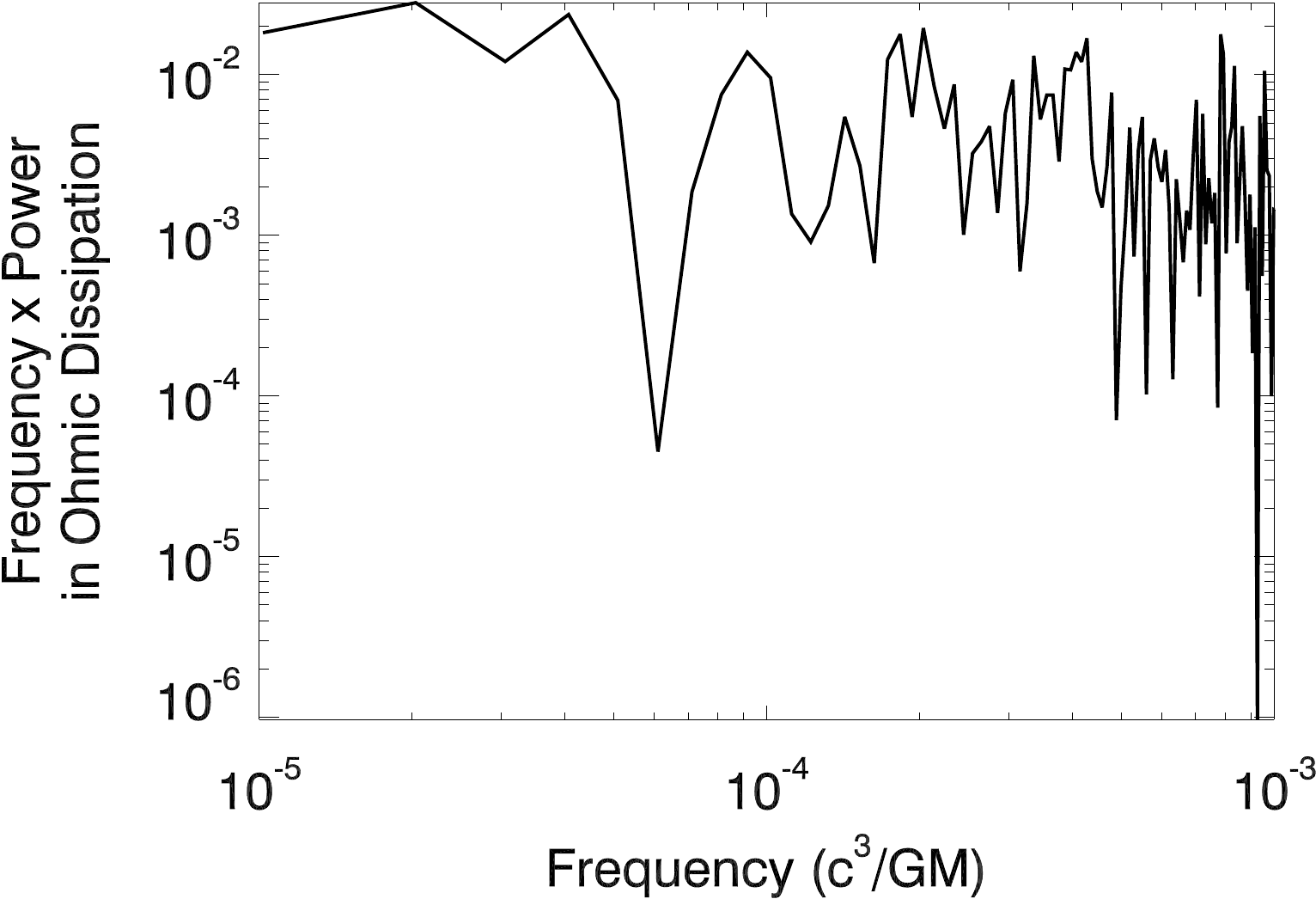}
\includegraphics[type=pdf,ext=.pdf,read=.pdf,height=0.2\textwidth]{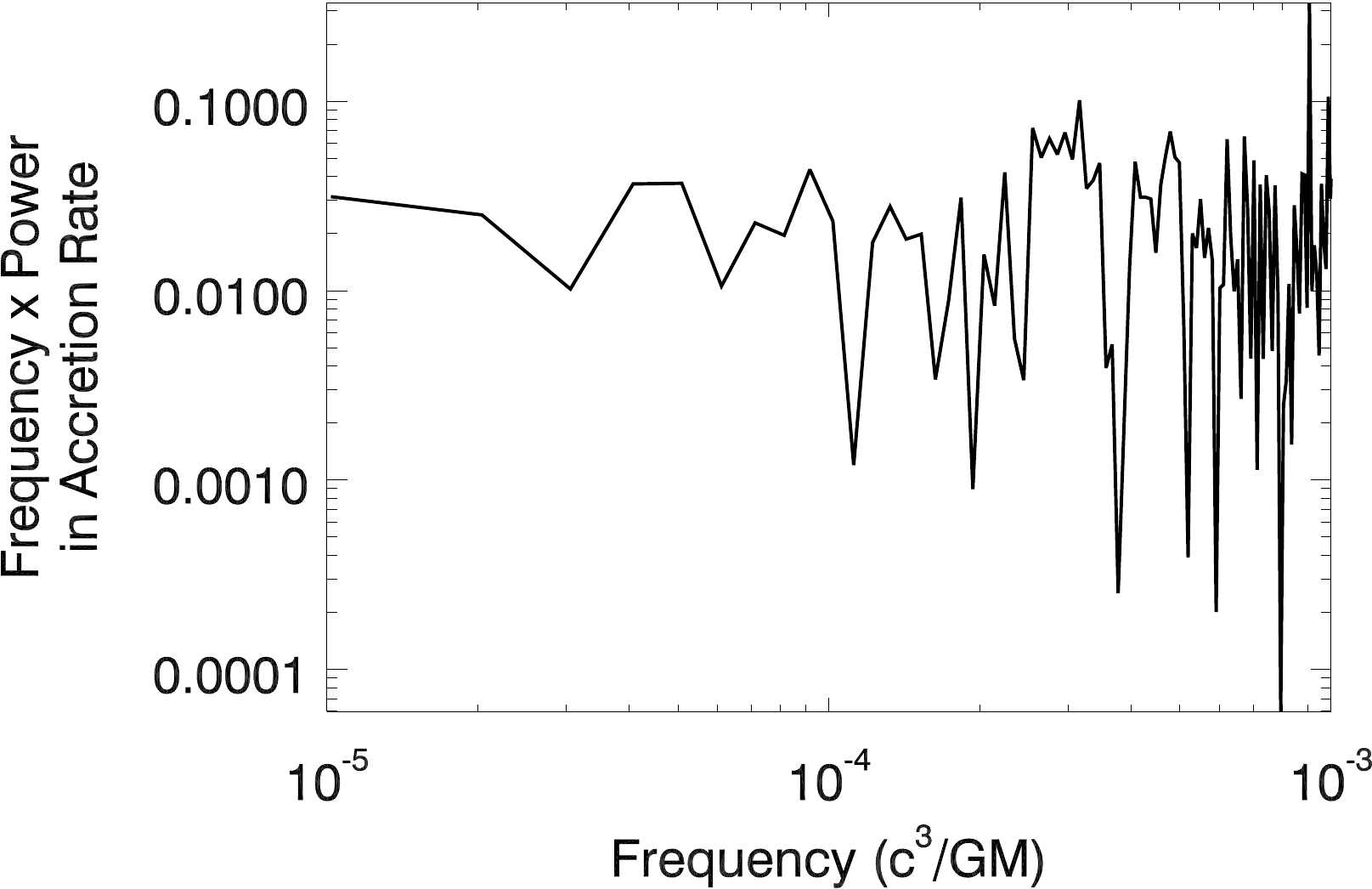}
\caption{Frequency-weighted PSDs ($\nu P$, where $P$ is RMS-normalized) of the total stresses ({\it left}), Ohmic dissipation ({\it center}), and accretion rate across the ISCO ({\it right}).  The stress and dissipation are integrated over a range that spans $R=15-25~r_{\rm g}$, stretching in elevation from the disk midplane to five scale heights above it.  The accretion rate is measured over a range that spans five scale heights above the midplane.  While the integrated stress shows signs of the azimuthal field signals seen in Figures 2-3, the dissipation and accretion rate are noise-dominated.}
\end{center}
\end{figure*}

\begin{figure*}[t]
\begin{center}
\includegraphics[type=pdf,ext=.pdf,read=.pdf,height=0.2\textwidth]{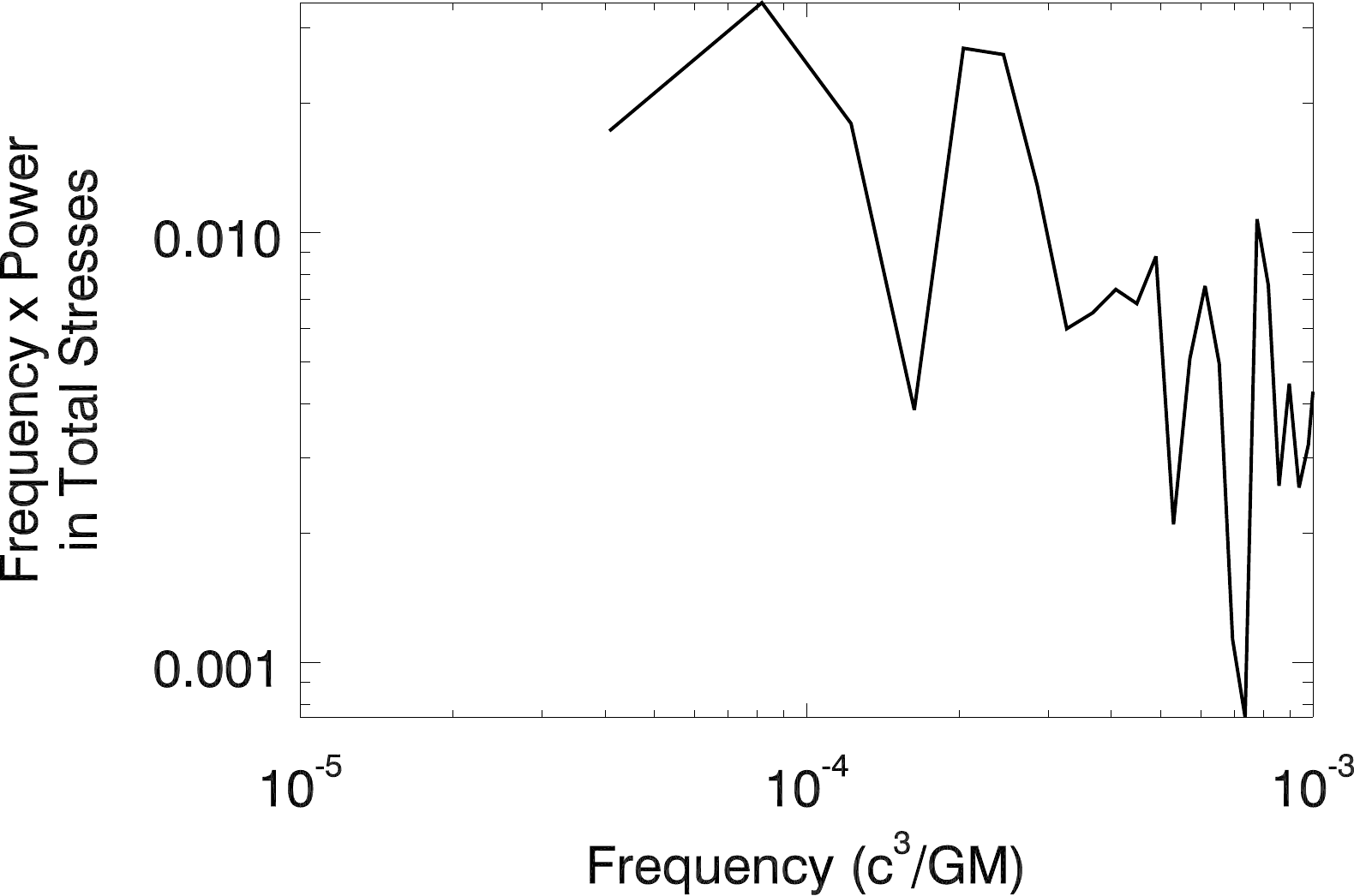}
\includegraphics[type=pdf,ext=.pdf,read=.pdf,height=0.2\textwidth]{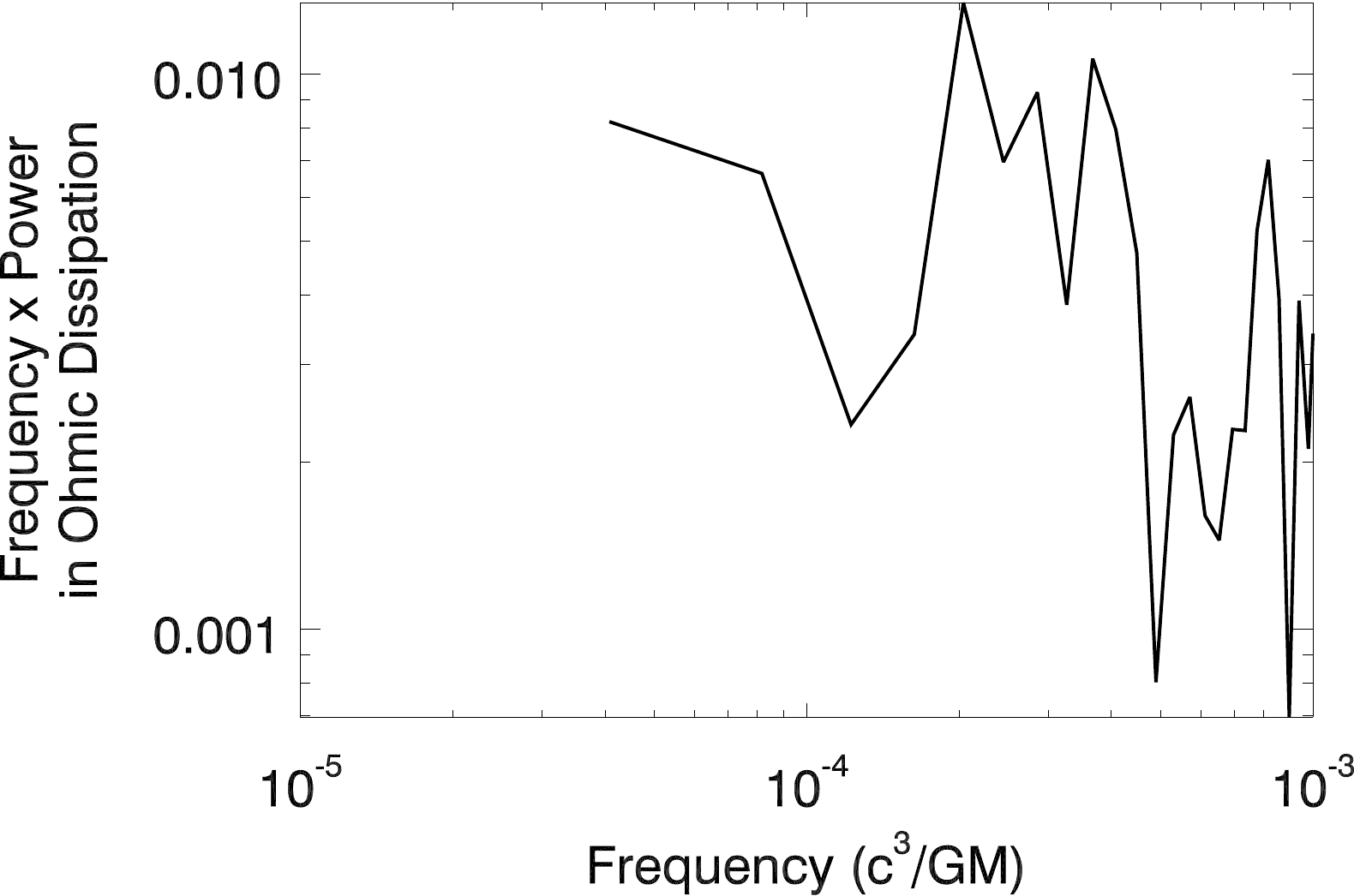}
\includegraphics[type=pdf,ext=.pdf,read=.pdf,height=0.2\textwidth]{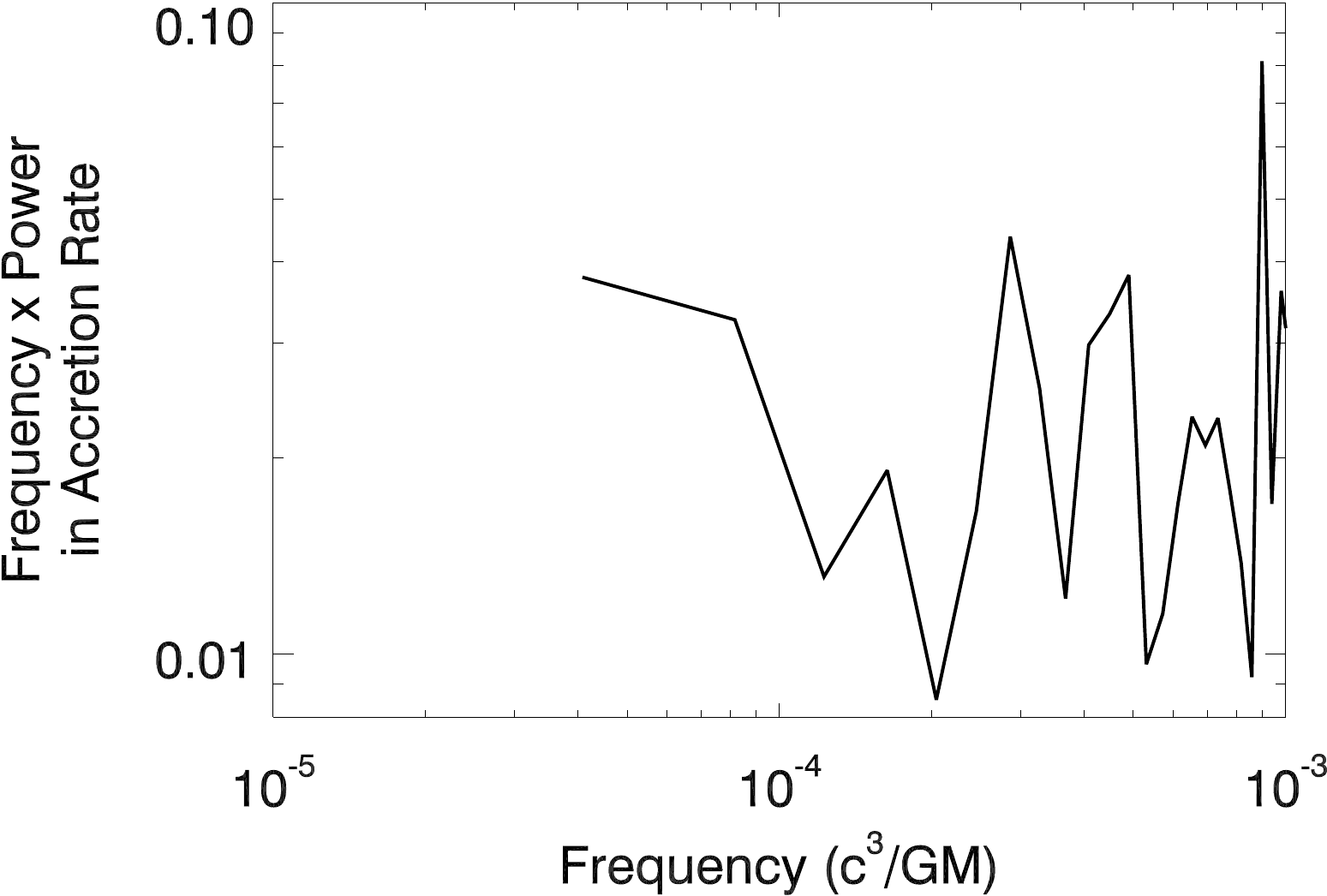}
\caption{Frequency-weighted $\overline{\rm PSD}$s ($\nu \overline{P}$, where $\overline{P}$ is RMS-normalized) of the total stresses ({\it left}), Ohmic dissipation ({\it center}), and accretion rate across the ISCO ({\it right}), as in Figure 6 but averaged over four independent time series.  While there are features in each panel near the frequencies of interest (\ie a few $\times 10^{-4} {\rm~c^3/GM}$), these features are not particularly well separated from the noise.}
\end{center}
\end{figure*}

As was done in Figure 3, we also construct the $\overline{\rm PSD}$s for these three observational proxies as averaged over four independent time series to evaluate the significance of the peaks in Figure 6.
For each proxy, we see features near the frequencies of interest (\ie a few $\times 10^{-4} {\rm~c^3/GM}$), but they are in all cases comparable (within approximately a factor of two) to the noise levels in the nearby lowest frequency bins, as seen in Figure 7.
This is in stark contrast to the case of the azimuthal magnetic field (Figures 2 and 3), where a clear signal stood above the noise in both the full PSDs and the $\overline{\rm PSD}$s constructed from the subdivided series.
Without better frequency resolution from a longer simulation, it is difficult to conclude that any peaks in the observational proxies are present at high levels of significance.

\section{Discussion}
\subsection{Interpretation and Comparison with Local Simulations of Accretion Disks}
We propose that the magnetic field oscillations clearly seen in Figures 1-5 are the manifestation of a dynamo cycle phenomenon.
As discussed in the Introduction, such cycles have been seen in local simulations of magnetized accretion that reflect a large variety of numerical algorithms and disk parameters.
Our work has shown that similar cycles are indeed present in global simulations of black hole accretion disks.
As in the local case, the azimuthal magnetic field cycles with frequencies approximately ten to twenty times lower than the local orbital frequency.
Individual peaks in these cycles can share power across large radial ranges, while the range of cycle frequencies remains roughly proportional to the local orbital frequency.
Additionally, these features persist for the full duration of the simulation (nearly $10^5$ GM/c$^3$).
This is much longer than the time expected to erase all memory of the initial field conditions (see \citealt{2010ApJ...712.1241S}, for example) and is comparable to the radial drift timescale at $r=10$ r$_{\rm g}$, suggesting that this behavior can be maintained even over a period during which the disk evolves significantly.

Linking the amplitudes and frequencies of the observed dynamo cycles to the parameters of our global disk simulation is, in practice, quite challenging.
For example, \citet{2002ApJ...579..359B} illustrate with a series of numerical experiments that the resulting dynamo cycle frequency depends on the time evolution of the effective magnetic diffusivity.
In a grid-based simulation such as ours, this will generally vary with both the grid resolution and flow details.
Additionally, our simulated disk does not maintain a perfect steady-state.
For example, the surface density at a given radius can vary by $50\%$ during the course of the simulation as accretion proceeds.
This variation is not necessarily monotonic, however, and is accompanied by variation in both the accretion rate and effective ``alpha'' parameter.
It is therefore difficult to associate the potential frequency drift seen in the first panel of Figure 5, for example, with any obvious trend in the disk, although a longer time baseline would potentially enable such an identification.  
As such, we take a phenomenological approach to the observed dynamo cycles and, having noted their similarity to a well-established aspect of local simulations, now discuss how they relate to astrophysical observables.

\subsection{Relevance to Astrophysical Black Holes}

The astronomical phenomena most obviously analogous to our simulated dynamo cycles are the LFQPOs detected in multiple black hole binary systems by X-ray satellites such as {\it Ginga} and the Rossi X-ray Timing Explorer.
As summarized in \citet{2006ARA&A..44...49R}, LFQPOs appear in the nonthermal X-ray spectrum and have frequencies that range from 0.1 to 30 Hz.
Although the peaks are often seen to migrate quite rapidly in frequency, a given peak typically has a quality factor (Q $\equiv \nu/\Delta \nu$) $\gtorder 10$.
Observed LFQPOs can also be very strong, exhibiting RMS amplitudes of tens of percents.

Our simulated dynamo cycles have three properties in common with observed LFQPOs.
First, they occur in the expected frequency range.  
To choose an example, an observed 4 Hz LFQPO for a 10 $\Msun$ black hole corresponds to a frequency in natural units of $\nu \sim 2 \times 10^{-4}$ GM/c$^3$, which is representative of the range that we see in the simulations.
Second, several of the simulated dynamo cycle peaks have high quality factors.
In practice, the duration of our simulation permits a maximum quality factor $Q_{\rm max} \approx 5$ at a frequency of $2 \times 10^{-4}$ GM/c$^3$, but this value is achieved at multiple radii in Figure 2, for example.
Naturally, the quality factors of the peaks in the time-averaged $\overline{\rm PSD}$s shown in Figure 3 are lower (specifically, of order unity) as a result of the reduced frequency resolution available in this mode of analysis and, potentially, the non-stationarity of the time series.
Third, the dynamo cycles are seen to occupy the magnetized, low-density region that has been traditionally identified with the hard X-ray emitting corona (\eg \citealt{2000ApJ...534..398M}).
Thus, there is no difficulty in linking dynamo cycles to the region where we think observed LFQPOs originate.
That noted, we should point out that our simulation does not include any treatment of realistic radiative processes.
Neither do any of our constructed observational proxies reflect the magnetic field cycles with a high degree of significance, nor do the proxies feature RMS amplitudes comparable to those seen in observed systems.
As such, we cannot make any detailed predictions concerning how these dynamo cycles would manifest themselves observationally.

Whatever the advantages and disadvantages of dynamo cycles as a model for LFQPOs, it is worth briefly distinguishing them from the most popular alternative models of LFQPO production.
First, that dynamo cycles are most prominent in the azimuthal magnetic field off the disk midplane shows that they have none of the obvious characteristics of trapped waves (\citealt{1990PASJ...42...99K}) or diskoseismic modes \citep{1991ApJ...378..656N,1992ApJ...393..697N,1993ApJ...418..187N,2009ApJ...692..869R,2009ApJ...693.1100O}, both of which should manifest themselves in the hydrodynamic variables.
Lense-Thirring precession has been explored in the context of LFQPOs by \citet{1996ApJ...458..508I} and \citet{1999ApJ...524L..63S}, for example, in the test particle limit and on a more global scale by \citet{2009MNRAS.397L.101I}, but our numerical model of a non-rotating black hole is obviously incapable of producing such effects.  
(In fact, given the observed dynamo cycles' locations and nature, one expects a similar outcome from a fully general relativistic simulation.)
The truncated disk model for LFQPOs proposed by \citet{2004A&A...427..251G} is also quite distinct from ours since dynamo cycles naturally produce oscillation frequencies much lower than the local orbital frequency even at tens of gravitational radii, requiring no disk cutoff. 
Finally, we note that dynamo cycles are not simply a subtle manifestation of the Accretion-Ejection Instability (AEI, \citealt{1999A&A...349.1003T}) since the AEI occurs near the inner edge of the disk while our cycles are seen to originate from all radii that have had sufficiently many orbital periods over which to evolve.

\section{Conclusions}
We have described a global, numerical, MHD study of black hole accretion that has revealed an interesting pattern of magnetic variability that, until now, had only been seen in local shearing box simulations.
The most important results of our study are summarized here:

1) We have identified for the first time the presence of dynamo cycles in global simulations of black hole accretion disks.
These cycles manifest themselves as oscillations in the azimuthal magnetic field in a region that stretches from a few to several scaleheights above the disk midplane in elevation.
Individual peaks in these cycles share power radially, while their frequency range at a given radius is found to be approximately ten to twenty times lower than the local orbital frequency.

2) While dynamo cycles are easily seen in the azimuthal magnetic field, detecting cyclic variation in derived quantities is much more challenging.
The integrated stresses feature variability that we have identified with the azimuthal field behavior, reflecting the fact that distinct peaks in the PSD are manifested by a wide range of radii.
The mass accretion rate and integrated Ohmic dissipation, however, remain noise-dominated, and none of the three proxies we have examined feature RMS amplitudes as large as those seen observationally.

3) We have discussed potential links between dynamo cycles and observed LFQPOs in black hole binaries.
While dynamo cycles naturally produce oscillations at the appropriate frequencies and locations expected for LFQPOs, any complete theory of LFQPOs will need to address how the field oscillations generate an observable signature, how their periodicity varies in time, and, ultimately, how low-frequency variability is connected to the inferred black hole accretion state.
Additionally, the duration of our simulation limits our LFQPO peaks to have quality factors Q $\leq 5$ when estimated from a single time series.
Such quality factors are still a factor of two below what is observed, implying that direct comparisons of this nature with observed LFQPOs will require extended simulations that run for several thousand orbits (or longer, if one wishes to average multiple PSDs together).
Nonetheless, our work shows that dynamo cycles in global simulations can produce oscillations with frequencies much lower than any natural frequency in the test particle regime. 

\acknowledgments
We acknowledge the support of NSF Grant AST 06-07428 (all authors), NASA ATP grants NNX10AE41G (C.S.R., K.A.S.) and NNX09AG02G (S.M.O.),  and the Maryland-Goddard Joint Space Science Institute (JSI) graduate fellowship program (K.A.S.).
We also thank Philip Cowperthwaite and Brett Morris for their work in visualizing these flows and the NCSA for developing ZEUS-MP.
This research was supported in part by the NSF through TeraGrid resources at the Texas Advanced Computing Center under grant TG-AST090105.
We also thank the anonymous referee for their comments.

\end{document}